\documentclass{aa}
\usepackage{graphicx}
\usepackage{amssymb}

\def\beqa{\begin{eqnarray}}
\def\eeqa{\end{eqnarray}}
\def\beq{\begin{equation}}
\def\eeq{\end{equation}}

\begin{document}

\thesaurus{2(12.07.1, 09.03.1, 08.03.4, 08.16.5, 08.16.4,
09.16.1)}

\titlerunning{Microlensing by Compact Objects associated to Gas Clouds}

\title{Microlensing by Compact Objects associated to Gas Clouds}

\authorrunning{V. Bozza et al.}

\author{V. Bozza$^{1,5}$\thanks{%
E-mail valboz@sa.infn.it}, Ph. Jetzer$^{2,3}$\thanks{%
E-mail jetzer@physik.unizh.ch}, L. Mancini$^{2,1,5}$\thanks{%
E-mail mancini@physik.unizh.ch}, G. Scarpetta$^{1,4,5}$\thanks{%
E-mail scarpetta@sa.infn.it}}

\institute{$^{1}$Dipartimento di Fisica ``E.R. Caianiello'',
Universit\'{a} di Salerno, I-84081 Baronissi (SA), Italy \\
$^{2}$Institut f\"{u}r Theoretische Physik der Universit\"{a}t
Z\"{u}rich, CH-8057 Z\"{u}rich, Switzerland \\ $^{3}$Institut
f\"ur Theoretische Physik der ETH, CH-8093 Z\"urich, Switzerland
\\ $^{4}$International Institute for Advanced Scientific Studies, Vietri sul Mare (SA), Italy \\
$^{5}$Istituto Nazionale di Fisica Nucleare, sez. Napoli, Gruppo
Collegato di Salerno, Italia}

\date{accepted/received}
\maketitle

\begin{abstract}
We investigate gravitational microlensing of point-like lenses
surrounded by diffuse gas clouds. Besides gravitational bending,
one must also consider refraction and absorption phenomena.
According to the cloud density, the light curves may suffer small
to large deviations from Paczy\'{n}ski curves, up to complete
eclipses. Moreover, the presence of the cloud endows this type of
microlensing events with a high chromaticity and absorption lines
recognizable by spectral analysis. It is possible that these
objects populate the halo of our galaxy, giving a conspicuous
contribution to the fraction of the baryonic dark matter. The
required features for the extension and the mass of the cloud to
provide appreciable signatures are also met by several
astrophysical objects. 
\end{abstract}

\section{Introduction}

In the last years \textit{\itshape{gravitational microlensing}}
has rapidly grown to an important astrophysical tool with wide
applications. In particular, it gave some evidence for the
presence of baryonic dark matter, in form of compact objects
(MACHOs), in the Galatic halo (Alcock et al. 2000a, Lasserre et
al. 2000) although the number of microlensing events in the
direction of the Magellanic Clouds, reported by the observational
teams MACHO and EROS, is lower by a factor $\sim5$ as predicted by
the standard halo models. The estimated MACHO fraction is now
about $20\%$ of the total Galactic dark matter mass, though it
must be kept in mind that this value is still affected by a very
high uncertainty. On the other hand several hundred of events have
been found towards the Galactic bulge region (Alcock et al. 2000b,
Wo\'{z}niak et al. 2001) and few events towards some spiral arm
regions (Derue et al. 2001).
Most likely the lenses are in these cases low mass stars
or brown dwarfs.

Classical microlensing consists of the magnification of the
luminous flux coming from a star due to the presence  of a
deflecting mass close to the line of sight. Within this picture,
microlensing yields an indirect observational method for the
detection of non-luminous bodies.

However, compact objects are not the only candidates for baryonic
dark matter in the halo. Some years ago, it has been proposed that
an appreciable fraction of the Galactic dark matter might be in
form of self-gravitating gas clouds with masses of the order of
Jupiter's mass ($10^{-3} M_{\odot}$) and radii $R \approx 10$ AU
(Pfenninger et al. 1994; De Paolis et al. 1995a, 1995b, 1996,
1998a, 1998b; Gerhard \& Silk 1996). For diffuse objects two more
phenomena must be taken into account. Traveling through a gaseous
medium, light basically undergoes refraction and absorption. So,
even if these gas clouds would produce no appreciable microlensing
effect, classical refraction could supply an alternative
amplification mechanism. The resulting light curves resemble
gravitational microlensing ones, but can be distinguished by their
peculiar features (Draine 1998; Rafikov \& Draine 2001).

In this paper, we mean to study an intermediate possibility, which
might be of relevance for the estimate of baryonic dark matter in
the halo. After almost ten years of microlensing campaigns, at a good
confidence level, we can say that MACHOs exist. If gas clouds are
effectively present in the halo, they could be likely associated
to MACHOs. If this is the case, these Pointlike Lenses Associated
to Gas (hereafter, PLAG) would be missed by standard microlensing
observations, because their light curve would be significantly
altered by the surrounding gas cloud. On the other hand, isolated
gas clouds could be quite difficult to detect without a powerful
amplification mechanism.

In PLAGs gravitational lensing by the central object would provide
this mechanism, but the secondary effects due to the cloud would
show up in a weak or strong fashion, depending on the density of
the diffuse matter. Classical absorption by Rayleigh scattering
would essentially block all light attempting to cross high density
clouds. For lower densities, a moderate absorption would be
contrasted by microlensing and refraction both concurring in the
amplification of the flux coming from the source. In this case, a
considerable reddening of the light must be expected, producing a
high chromaticity in the light curve. Of course, the existence of
the gas cloud could be revealed without doubts by taking a
spectrum of the source during the microlensing event. In this way,
the chemical composition of the matter surrounding the central
compact object can be determined.

Several studies about the stability of gas clouds seem to favour
the formation of a steady state in protostellar nebulae (Cazes \&
Tohline 1998). If MACHOs were formed by the collapse of gas
clouds, a large portion of this cold gas may still surround some
of them, with a reasonable formation of PLAGs.

The abundance of dark PLAGs in the halo is not known, leaving open
several possibilities. They could be completely absent, if some
instability mechanism intervenes. They may be a negligible
fraction of MACHOs, if their mean life is too small. On the
contrary, they could be relatively abundant and partly explain the
discrepancy between observations and theoretical predictions about
the microlensing rate, since the missing microlensing events would
be actually PLAG events, not detected by usual observations. If
this is true, then microlensing experiences should be properly
modified in order to detect PLAG events efficiently.

Besides dark PLAGs in the halo, many well-known astrophysical
compact objects surrounded by gas are present in the galactic
disk. We can mention, for instance, planetary and proto-planetary
nebulae, proto-stars, LBV nebulae, nebular variables (T Tau stars
and R W Aur stars), eruptive variables (R CrB stars, Wolf-Rayet
stars, novae), and so on. Of course, not all of them are able to
produce microlensing effects that are really observable by
monitoring programs, since their parameters should satisfy several
constraints.

A fundamental tuning is required between the size of the cloud
$R_c$ and the Einstein radius $R_{E}$. The ratio $R_c/R_E$ should
be in the range $10^{-1}\div 10^2$. This tuning is demanded by the
necessity of having a microlensing time comparable to the time
required by the source to pass behind the whole gas cloud. In
fact, if the cloud is too small, it would act, at all effects, as
a compact object. On the contrary, if the cloud is much larger
than the Einstein radius, with the typical velocities of galactic
objects, the time taken for the source to pass behind the whole
cloud may easily rise up to years. However, even in this case,
these long duration events could be included among the objectives
of observational campaigns following stars for very long periods.

For $R_c/R_E\gtrsim 10^2$, we would observe no deviation from the
usual microlensing light curve, since the effects of the cloud
would be manifest after decades and could not be practically
noticed. No information on the size and the shape of the cloud
would be retrieved in this case.

In Sect. 2, we introduce the basics of the three phenomena
affecting the light passing through the cloud: gravitational
lensing, refraction and absorption by Rayleigh scattering. In
Sect. 3, we analyze the general features of the light curves
expected from PLAGs. In Sect. 4, we briefly discuss the properties
of the afore-mentioned PLAG candidates. Sect. 5 contains the
summary.

\section{Light passing through gas clouds: basic phenomena}

We restrict our analysis to spherical gas clouds surrounding
point-like masses. This simplification allows an extensive
analytical investigation. Moreover, spherically symmetric PLAGs
should be easier to distinguish from variable stars and other
usual background to microlensing.

The lens equation contains both refraction and gravitational
lensing. The latter can be separated into two contributions: the
deviation due to the central object and the one produced by the
gas cloud. Thanks to the spherical symmetry, we can write the one
dimensional equation

\begin{equation}
\label{lens equation}
y=x-\frac{D_{ol}D_{ls}}{D_{os}}\left[\alpha_{0}(x)+\alpha_{c}(x)+\alpha_{r}(x)\right],
\end{equation}
where
\begin{equation}
\alpha_{0}(x)=\frac{4 G M}{c^{2}x} \label{alpha_0}
\end{equation}
is the deflection angle produced by the gravitational field of the
compact object (of mass $M$) at the center of the PLAG (Schneider,
Ehlers and Falco 1992); $\alpha_c(x)$ is the deviation angle due
to the gravitational field of the gas cloud; $\alpha_r(x)$ is the
contribution to the deviation coming from refraction of light
passing through the cloud.  As usual, $y$ ($x$, respectively) is
the distance of the source (of the image, respectively) to the
optical axis, defined as the line connecting the observer and the
lens.

The splitting of the deviation angle into the sum of three
separate contributions is possible in the linear approximation.
Higher orders would mix up the three effects.

The gas cloud is described by a continuous density distribution
$\rho(r)$ which is a function of the radial distance from the center
of the PLAG, according to the spherical symmetry hypothesis. Since
the PLAG has limited size, we call $R_c$ the value of the radius
where the density falls to zero. The total mass of the cloud is
$M_c$. The projected density is
\begin{equation}
\Sigma(x)=\int_{-\sqrt{R_c^2-x^2}}^{\sqrt{R_c^2-x^2}}
\rho(\sqrt{x^2+z^2}) dz.
\end{equation}

The gravitational bending for light rays passing outside the cloud
($x>R_c$) is just $\frac{4 G M_c}{c^{2}x}$, while for $x<R_c$,
$\alpha_c$ decreases to zero. It is generally given by
\begin{equation}\label{alpha_c}
\alpha_{c}(x)= \frac{4 G}{c^{2}}\frac{2\pi}{x} \int_{0}^{x} \Sigma
\left(x' \right)
 x' dx'=\frac{4G m(x)}{c^2x}~,
\end{equation}
where $m(x)$ is the mass of the portion of the cloud contained in
the cylinder of radius $x$ (Schneider, Ehlers and Falco 1992).

Light rays passing inside the gas cloud undergo Rayleigh
scattering by neutral gas (essentially Hydrogenum and Helium) and
eventually dust. The macroscopic consequences are refraction and
extinction, both depending on the density of the neutral gas
$\rho_n(r)$ and on its chemical composition.

The refraction index of a gas is
\begin{equation}
n_\lambda (r)=1+\alpha_\lambda \rho_n(r),
\end{equation}
where $\alpha_\lambda$ is the specific refractivity. The deviation
angle can be easily derived by the Fermat principle assuming the
deviation to be small. Its expression is
\begin{equation}
\label{alpha_r} \alpha_{r}(x)=-\alpha_\lambda
\frac{d\Sigma_n}{dx}=-2 \alpha_\lambda x \int_{x}^{R_c}
\frac{dr}{\sqrt{r^{2}-x^{2}}} \frac{d\rho_n}{dr}.
\label{Refraction angle}
\end{equation}
where $\Sigma_n$ is the projected density of neutral gas (confront Draine 1998,
Eq.1).

The magnification of the images follows from the one dimensional
lens equation (\ref{lens equation})
\begin{equation}\label{general magnification}
\mu=\frac{x}{y\frac{dy}{dx}}
\end{equation}
with y given by the RHS of Eq. (\ref{lens equation}). This
magnification factor is a consequence of the geometrical bending
on the light rays, caused by refraction and gravitational field.

The observed intensity is moreover modified by combined effects of
scattering and absorption, giving rise to the so-called
extinction, described by the Lambert's law

\begin{equation}
\label{transfer equation 2} I_{\lambda}=I_{\lambda,0}
e^{-\tau_{\lambda}},
\end{equation}
where $I_{\lambda,0}$ is the intensity that we see in absence of
absorption along the line of sight, i.e. before and after the
microlensing event, and $\tau_{\lambda}$ is the optical depth of
the cloud, given by
\begin{equation}
\tau_\lambda=k_\lambda\Sigma_n (x).
\end{equation}
where $k_{\lambda}$ is the absorption coefficient.

Summing up, each image will have a total amplification given by
the geometrical magnification (\ref{general magnification}) due to
light bending (both gravitational and refractive) weighted by the
transmission coefficient $e^{-\tau_\lambda}$.

\section{General properties of PLAGs}

The fundamental parameters of the PLAG are the mass of the central
object $M$, the radius of the cloud $R_c$ and its mass $M_c$. Most
statements about the aspect of the light curves can be given in
terms of these parameters, or their derived quantities, without
need to specify the shape of the density profile (if it is smooth
enough).

We define the Einstein radius
\begin{equation}
R_E=\sqrt{\frac{4G(M+M_c)}{c^2}\frac{D_{OL}D_{LS}}{D_{OS}}}
\end{equation}
referring to the total mass of the system.

In this section, using a simplified model, we will discuss the
phenomenology that should be generally expected in microlensing
events caused by PLAGs.

\subsection{Sample family of PLAGs}

If we consider PLAGs formed by hot nebulae, in the central region
the gas maybe ionized, being uneffective for classical Rayleigh
scattering, suggesting $\rho_n < \rho$. We are mainly interested
in cold dark PLAGs; so, in a first approach, we will assume that
this region, if present, is small enough with respect to the total
cloud and the Einstein radius, so that $\rho_n\simeq \rho$. In
this way, we can rapidly obtain a simple description of the basic
phenomenology. Nevertheless, the extension to the case $\rho_n <
\rho$ is straightforward and can be easily faced successively.

The dependence on the nature of the gas is not so strong to
substantially affect the discussion, since the order of magnitude
of the specific refractivity and the absorption coefficient is the
same for all the interesting elements or molecules. Thus, in all
practical examples, we will assume a mixture of H$_{2}$-He with
$24\%$ He by mass. At $\lambda= 4400$ \AA \, for this mixture, we
have (AIP Handbook 1972, following Draine 1998)
\begin{eqnarray}
\alpha_\lambda&\simeq&1.243 cm^{3} g^{-1} \\
k_\lambda &\simeq& 5.4 \times 10^{-5} \left( \frac{0.44 \mu m
}{\lambda}\right)^4 cm^2 g^{-1} \label{absorption coefficent}
\end{eqnarray}

The variety of density profiles in nebulae is very large and most
of them have no satisfying description at the present time. This
variety seems to make the choice of the density model a hard task.
However, once projected along the line of sight, all realistic
density distributions are generally smoothly peaked at the center
and decay towards the border in a more or less steep fashion. For
this reason, the basic phenomenology of PLAG events does not
depend on the specific shape of the original density profile but
can be discussed by analyzing one representative model.

We choose the function

\begin{equation}
\rho_\beta(r)= \left\{
  \begin{array}{ll}
      \rho^0_\beta \left(1-\frac{r^2}{R_c^2} \right)^\beta & \mbox{for $r \leq R_{c}$} \\
       0 & \mbox{otherwise}
      \end{array} \right. ,
      \label{rho_beta}
\end{equation}
parameterized by the exponent $\beta>0$, to describe the density
profile of a typical PLAG. The choice of this model is strongly
motivated by the fact that most physically interesting quantities
can be calculated analytically starting from this distribution,
while all the general features of a PLAG microlensing event can be
obtained. Moreover, the parameter $\beta$ can be adjusted to suit
to distinct physical situations: high $\beta$'s would produce
clouds concentrated at the center with smooth boundaries which
would appropriately describe stable or nearly stable systems; low
$\beta$'s, on the contrary, can be used for expanding clouds,
which are generally characterized by a steep boundary and where
abrupt changes in the density may produce discontinuities in the
refraction angle (see Sect. 3.4 and Fig. \ref{Fig deviation
angles}b).

Of course, if one is interested in the details of the microlensing
light curves, it is necessary to know the real shape of the
nebula; but for all the highlights of microlensing light curve,
the study of this sample model will be sufficient.

The central density can be expressed in terms of the mass of the
cloud
\begin{equation}
M_c=\int_0^{R_c} 4\pi r^2 \rho_\beta(r)dr= \rho_\beta^0
\frac{\pi^{3/2} R_c^3 \Gamma\left[\beta+1
\right]}{\Gamma\left[\beta+\frac{5}{2} \right]}.
\end{equation}

The projected density is
\begin{equation}
\Sigma_\beta(x)=\rho_\beta^0\frac{\pi^{1/2} R_c
\Gamma\left[\beta+1 \right]}{\Gamma\left[\beta+\frac{3}{2}
\right]}\left(1-\frac{x^2}{R_c^2} \right)^{\beta+\frac{1}{2}}.
\end{equation}

The gravitational bending due to this density distribution is
\begin{equation}
\alpha_c(x)=\frac{4G\rho_\beta^0 \pi^{3/2} R_c^3
\Gamma\left[\beta+1 \right]}{c^2 x\Gamma\left[\beta+\frac{5}{2}
\right]}\left[1-\left(1-\frac{x^2}{R_c^2}
\right)^{\beta+\frac{3}{2}} \right].
\end{equation}

The refraction angle is
\begin{equation}
\alpha_r(x)=\alpha_\lambda \frac{2 \beta \rho_\beta^0 \pi^{1/2}
R_c^3 \Gamma\left[\beta \right]}{\Gamma\left[\beta+\frac{1}{2}
\right]}\frac{x}{R_c} \left(1-\frac{x^2}{R_c^2}
\right)^{\beta-\frac{1}{2}}.
\end{equation}

Of course, we cannot take as physical the divergence for
$0<\beta<1/2$. The simple expression we are using for the
refraction angle is valid for small deviations, so the presence of
such divergence signals a breakdown of this hypothesis and the
need for higher order corrections. These corrections should
provide a cutoff for the refraction angle.

\begin{figure}
\resizebox{\hsize}{!}{\includegraphics{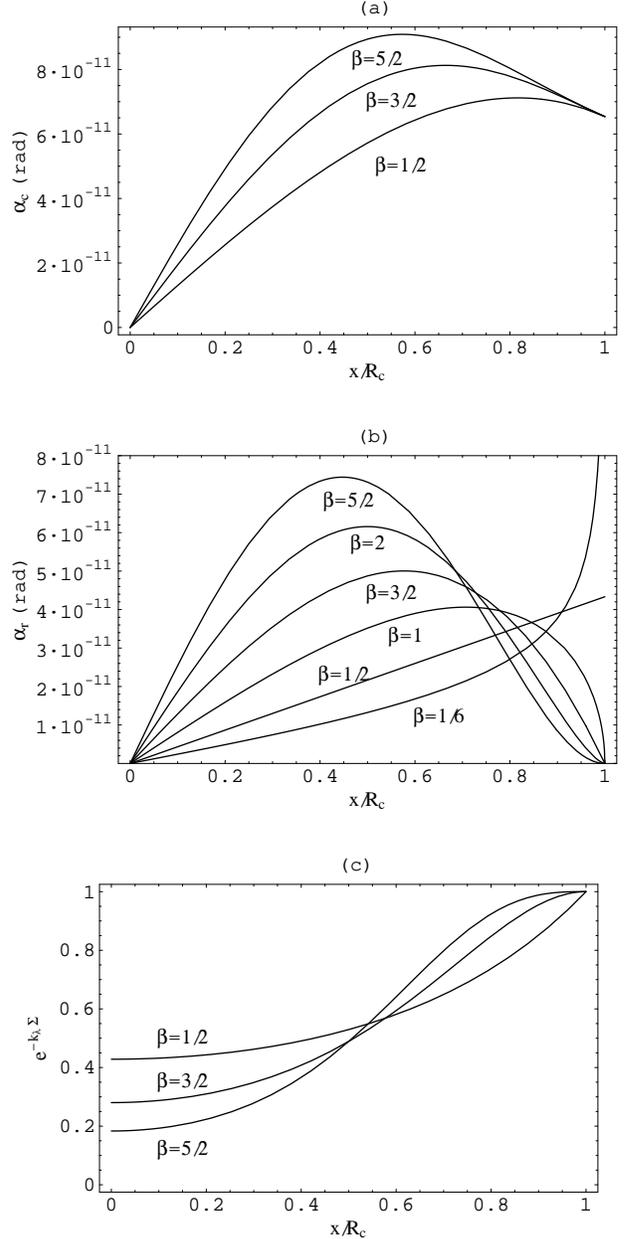}}
 \caption{(a) Gravitational bending due to the cloud. (b) Refraction angle.
 (c) Transmission coefficient. All quantities are plotted for different
 values of $\beta$.}
 \label{Fig deviation angles}
\end{figure}

In Fig. \ref{Fig deviation angles} we plot the gravitational
bending, the refraction angle and the transmission coefficient of
our cloud model for $M_{c}=0.01 \, M_{\odot}$, $R_{c}=6$ AU and
for different values of $\beta$.

\subsection{Light curves}

\begin{figure*}
\resizebox{15cm}{!}{\includegraphics{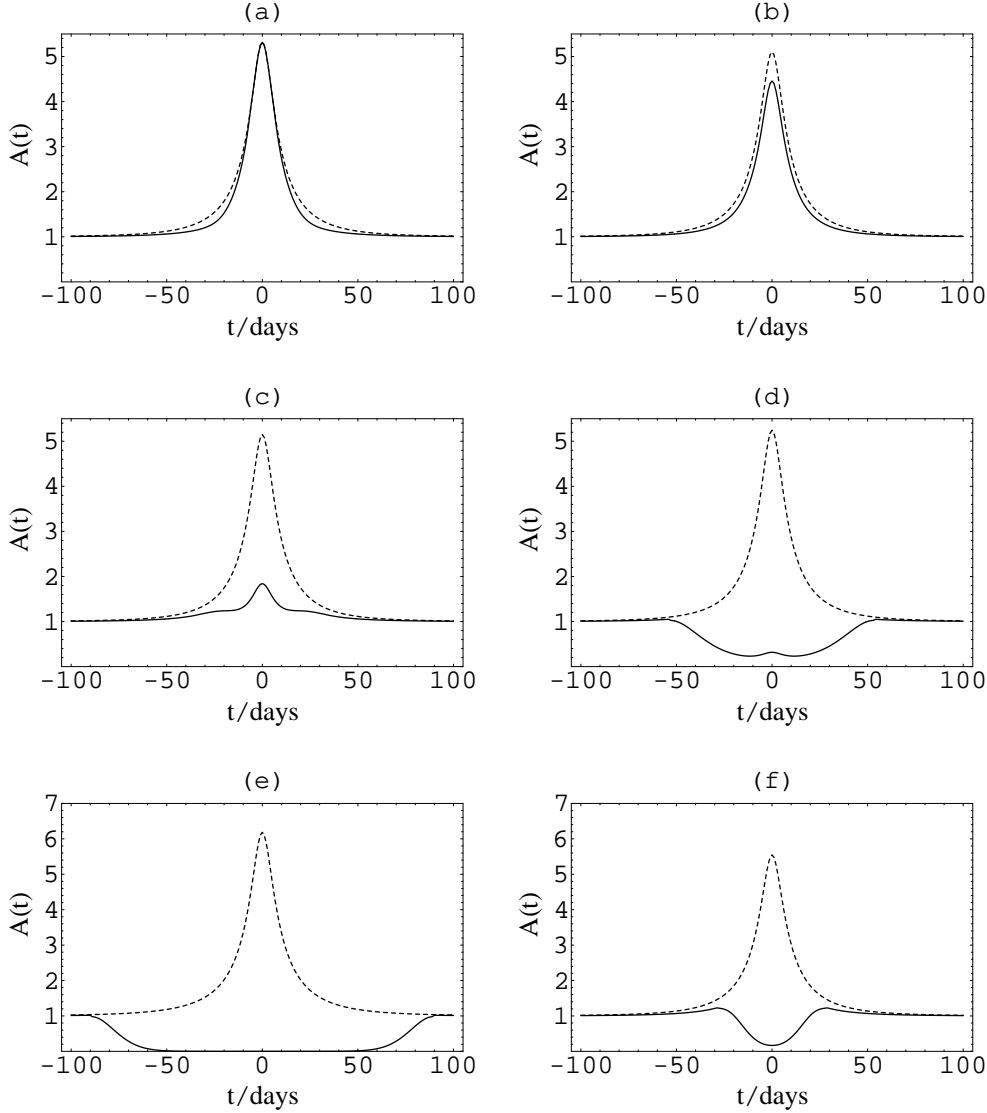}}
 \caption{The six categories of light curves expected for PLAG events
as discussed in Sec.3.2. The dotted lines represent the classical
Paczy\'{n}ski light curves for gravitational lensing by the total
mass $M+M_c$. The common parameters, typical for a disk/bulge
microlensing events, are $\beta=5/2$, $D_{ol}=8 \, kpc$, $D_{ls}=2
\, kpc$, $M=1\ M_{\odot}$, impact parameter $b=0.2 R_E$ and
transverse velocity $v_T=200 km/s$. The other parameters are: (a)
$M_{c}=0.01 M_{\odot}$, $R_{c}=3.3$ AU; (b) $M_{c}=0.01
M_{\odot}$, $R_{c}=4.5$ AU; (c) $M_{c}=0.03  M_{\odot}$, $R_{c}=6$
AU; (d) $M_{c}=0.7 M_{\odot}$, $R_{c}=8$ AU; (e) $M_{c}=0.5
M_{\odot}$, $R_{c}=12$ AU; (f) $M_{c}=0.2  M_{\odot}$, $R_{c}=6$
AU.}
 \label{Fig Sample Curves}
\end{figure*}

We can distinguish two main classes of PLAGs, according to the
ratio between $R_c$ and $R_E$.

In small PLAGs ($R_c<R_E$), the positive parity image (the image
formed outside of the Einstein ring) is not affected by the
existence of the gas cloud. Its position is just that of the
principal image of a point-like lens with mass $M+M_c$. Since
light passes outside of the cloud, neither refraction nor
absorption modify the simple gravitational lensing effect.

The secondary image has negative parity and is formed inside the
Einstein ring. If the source is close enough to the optical axis,
this image is external to the cloud as well. In this situation,
the PLAG acts as a point-like lens at all effects. When the source
is far from the optical axis, the secondary image is formed inside
the PLAG. In this case, its position is modified by the non
trivial form of the deviation angles $\alpha_r$ and $\alpha_c$.
Moreover, the light flux coming from this image is partially
absorbed. However, the secondary image is relevant for the total
flux only when it is close to the Einstein ring, otherwise it is
highly demagnified. So, these additional effects intervene in a
regime where the image does not contribute much to the light
curve. On the basis of these considerations, the typical light
curve we expect for $R_c<R_E$ is a Paczy\'{n}ski-like curve
slightly lowered far from the maximum, where the secondary image
enters the cloud. This deviation can be easily confused in the
background photometric noise.

To ensure the effect to be appreciable, the size of the cloud
should be as close as possible to the Einstein ring and the
density as large as possible. In fact, in this case, the secondary
image would remain inside the cloud during a large fraction of the
microlensing event. A large absorption would depress it until it
exits from the cloud. So, just the region in the neighborhood of
the peak would be pure microlensing, while outside of this
interval, the secondary image would be depressed, changing the
shape of the curve. An example is given in Fig. \ref{Fig Sample
Curves}a, drawn (like the others) using our sample model.

\begin{table*}
\centering
\begin{tabular}[h]{l|ccccccc}
   \hline
   \hline
    & $\frac{R_c}{R_E}=0.5$ & $\frac{R_c}{R_E}=0.75$ & $\frac{R_c}{R_E}=1$ & $\frac{R_c}{R_E}=1.25$ & $\frac{R_c}{R_E}=1.5$
   & $\frac{R_c}{R_E}=1.75$ & $\frac{R_c}{R_E}=2$ \\
   \hline
   $M_c/M=10^{-4}$ &  $4.9\times 10^{-5}$ & $4.9\times 10^{-5}$ & $5\times 10^{-5}$ & $6\times 10^{-5}$ & $7.2\times
   10^{-5}$ & $7.6\times 10^{-5}$ & $7.4\times 10^{-5}$ \\
   $M_c/M=10^{-3.5}$ & $1.5\times 10^{-4}$ & $1.5\times 10^{-4}$ & $1.6\times
   10^{-4}$& $1.9\times 10^{-4}$ & $2.3\times 10^{-4}$ & $2.4\times 10^{-4}$ &$2.4\times 10^{-4}$ \\
   $M_c/M=10^{-3}$  & $4.8\times 10^{-4}$ & $4.8\times 10^{-4}$
   & $5.1\times10^{-4}$ & $6.5\times 10^{-4}$ & $7.7\times 10^{-4}$ & $8\times 10^{-4}$ & $7.6\times 10^{-4}$\\
   $M_c/M=10^{-2.5}$ & $1.5\times 10^{-3}$&$1.5\times 10^{-3}$&$1.8\times
   10^{-3}$&$2.4\times 10^{-3}$&$2.9\times 10^{-3}$&$2.9\times 10^{-3}$&
   $2.6\times 10^{-3}$\\
   $M_c/M=10^{-2}$&$4.8\times 10^{-3}$&$5\times 10^{-3}$&$7.4\times 10^{-3}$&
   $1.1\times 10^{-2}$&$1.3\times 10^{-2}$&$1.2\times 10^{-2}$&$10^{-2}$\\
   $M_c/M=10^{-1.5}$&$1.5\times 10^{-2}$&$2.2\times 10^{-2}$&$3.3\times 10^{-2}$&
   $5.8\times 10^{-2}$&$7.4\times 10^{-2}$&$6.6\times 10^{-2}$&$5.2\times
   10^{-2}$\\
   $M_c/M=10^{-1}$&$4.5\times10^{-2}$&$9.6\times10^{-2}$& $0.13$&
   $0.28$&$0.4$&$0.37$&$0.29$\\
   $M_c/M=10^{-0.5}$&$0.12$&$0.26$&$0.42$&$0.83$&$0.95$&$0.95$&$0.90$\\
\hline
\end{tabular}
  \caption{Maximal fractional contribution to the complete light curve
  coming from the gravitational bending due to the cloud.}
  \label{Tab grav}
\end{table*}

\begin{table*}
\centering
\begin{tabular}[h]{l|ccccccc}
   \hline
   \hline
    & $\frac{R_c}{R_E}=0.5$ & $\frac{R_c}{R_E}=0.75$ & $\frac{R_c}{R_E}=1$ & $\frac{R_c}{R_E}=1.25$ & $\frac{R_c}{R_E}=1.5$
   & $\frac{R_c}{R_E}=1.75$ & $\frac{R_c}{R_E}=2$ \\
   \hline
   $M_c/M=10^{-4}$ &  $5.7\times 10^{-5}$ & $9.3\times 10^{-5}$ & $9.9\times 10^{-5}$ & $10^{-4}$ & $3.8\times
   10^{-5}$ & $2.9\times 10^{-5}$ & $2.9\times 10^{-5}$ \\
   $M_c/M=10^{-3.5}$ & $1.8\times 10^{-4}$ & $2.9\times 10^{-4}$ & $3.1\times
   10^{-4}$& $3.3\times 10^{-4}$ & $1.2\times 10^{-4}$ & $9.3\times 10^{-5}$ &$9.3\times 10^{-5}$ \\
   $M_c/M=10^{-3}$ &$5.5\times 10^{-4}$ & $9.1\times 10^{-4}$ & $9.8\times 10^{-4}$
   & $10^{-3}$ &$3.8\times 10^{-4}$ & $3\times 10^{-4}$ & $3.1\times 10^{-4}$ \\
   $M_c/M=10^{-2.5}$ & $1.6\times 10^{-3}$&$2.8\times 10^{-3}$&$3\times
   10^{-3}$&$3\times 10^{-3}$&$1.2\times 10^{-3}$&$10^{-3}$&
   $1.1\times 10^{-3}$\\
   $M_c/M=10^{-2}$&$4.3\times 10^{-3}$&$7.9\times 10^{-3}$&$8.9\times 10^{-3}$&
   $7.3\times 10^{-3}$&$4.3\times 10^{-3}$&$5\times 10^{-3}$&$4.5\times 10^{-3}$\\
   $M_c/M=10^{-1.5}$&$9.5\times 10^{-3}$&$2\times 10^{-2}$&$2.3\times 10^{-2}$&
   $1.3\times 10^{-2}$&$2.8\times 10^{-2}$&$3.3\times 10^{-2}$&$2.5\times
   10^{-2}$\\
   $M_c/M=10^{-1}$&$1.6\times 10^{-2}$&$4.3\times 10^{-2}$&$4\times 10^{-2}$&
   $6.4\times 10^{-2}$&$0.2$&$0.22$&$0.16$\\
   $M_c/M=10^{-0.5}$&$2.5\times 10^{-2}$&$7.1\times 10^{-2}$&$7.4\times
   10^{-2}$&$0.18$&$0.76$&$0.82$&$0.7$\\
\hline
\end{tabular}
  \caption{Maximal fractional contribution to the complete light curve
  coming from refraction.}
  \label{Tab refr}
\end{table*}

\begin{table*}
\centering
\begin{tabular}[h]{l|ccccccc}
   \hline
   \hline
    & $\frac{R_c}{R_E}=0.5$ & $\frac{R_c}{R_E}=0.75$ & $\frac{R_c}{R_E}=1$ & $\frac{R_c}{R_E}=1.25$ & $\frac{R_c}{R_E}=1.5$
   & $\frac{R_c}{R_E}=1.75$ & $\frac{R_c}{R_E}=2$ \\
   \hline
   $M_c/M=10^{-4}$ &  $3.8\times 10^{-4}$ & $8.2\times 10^{-4}$ & $1.3\times 10^{-3}$ & $1.8\times 10^{-3}$ & $3.2\times
   10^{-3}$ & $4.2\times 10^{-3}$ & $4.5\times 10^{-3}$ \\
   $M_c/M=10^{-3.5}$ & $1.1\times 10^{-3}$ & $2.6\times 10^{-3}$ & $4.2\times
   10^{-3}$& $5.7\times 10^{-3}$ & $10^{-2}$ & $1.3\times 10^{-2}$ &$1.4\times 10^{-2}$ \\
   $M_c/M=10^{-3}$ &$3.2\times 10^{-3}$ & $7.6\times 10^{-3}$ & $1.3\times 10^{-2}$
   & $1.7\times 10^{-2}$ &$3.1\times 10^{-2}$ & $4.1\times 10^{-2}$ & $4.4\times 10^{-2}$ \\
   $M_c/M=10^{-2.5}$ & $7.4\times 10^{-3}$&$2\times 10^{-2}$&$3.6\times
   10^{-2}$&$5.1\times 10^{-2}$&$9.5\times 10^{-2}$&$0.12$&
   $0.13$\\
   $M_c/M=10^{-2}$&$1.2\times 10^{-2}$&$4.3\times 10^{-2}$&$8.6\times 10^{-2}$&
   $0.13$&$0.26$&$0.34$&$0.36$\\
   $M_c/M=10^{-1.5}$&$1.3\times 10^{-2}$&$6.4\times 10^{-2}$&$0.15$&
   $0.29$&$0.57$&$0.7$&$0.74$\\
   $M_c/M=10^{-1}$&$9.9\times 10^{-3}$&$6.5\times 10^{-2}$&$0.19$&
   $0.49$&$0.86$&$0.96$&$0.98$\\
   $M_c/M=10^{-0.5}$&$6.1\times 10^{-3}$&$5\times 10^{-2}$&$0.18$&
   $0.62$&$0.98$&$0.9991$&$0.9999$\\
\hline
\end{tabular}
  \caption{Maximal fractional absorption.}
  \label{Tab abs}
\end{table*}

The time of the entrance of the secondary image into the cloud is
\begin{equation}
t=t_0+\frac{1}{v_T}\sqrt{\left(\frac{R_E^2}{R_c}-R_c \right)-b^2}
\label{Entrance time}
\end{equation}
where $b$ is the impact parameter of the source to the optical
axis, $v_T$ is the relative projected velocity and $t_0$ is the
time of the closest approach.

An interesting phenomenology arises when $R_c>R_E$. The secondary
image is always inside the cloud, being displaced and absorbed.
The principal image would enter the cloud only in a neighborhood
of $t_0$, for sufficiently small impact parameters. The time of
the exit of the principal image from the cloud is still given by
Eq. (\ref{Entrance time}). Depending on the density of the cloud,
we can distinguish some cases.

If the density is low, the absorption slightly lowers the maximum
with respect to Paczy\'{n}ski (Fig. \ref{Fig Sample Curves}b).

For moderate densities, the curve will be depressed when the
principal image enters the cloud. So, during the ascent to the
maximum, the curve will suddenly decrease its derivative and then
complete its rise to a considerably smaller maximum (Fig. \ref{Fig
Sample Curves}c).

For high densities, the absorption will prevail on the
magnification. When the principal image enters the cloud, the
light reaching the observer falls rapidly. Eventually, the
magnification at $t=t_0$ can still produce a low central peak in
the valley (Fig. \ref{Fig Sample Curves}d), but for higher
densities, even this faint feature would disappear. If the time of
the entrance is far from $t_0$, the event resolves in a complete
eclipse of the source (Fig. \ref{Fig Sample Curves}e); otherwise,
the initial part of the ascent can signal the presence of the
central lens inside the cloud (Fig. \ref{Fig Sample Curves}f).

The curves just described represent what can be generally expected
from this physical situation. As already quoted, the specific
shape obviously depends on the density profile considered. This
fact can be exploited to try a reconstruction of the density
profile from the details of the microlensing light curve.

We can make some simple considerations about the relative weight
of gravitational lensing, refraction and absorption in the light
curves, $A(t)$. Keeping $\beta=5/2$, we have compared the partial light
curves (obtained neglecting one of the above-quoted phenomena at a
time) with the exact ones. First, for each value of $M_c/M$ and
$R_c/R_E$, we have calculated the fractional deviation of the
curve without cloud gravitational lensing with respect to the
complete one, in the course of the whole microlensing event
\begin{equation}
f_{\alpha_c=0}(t)=\frac{\left| A(t)-A(t)|_{\alpha_c=0}
\right|}{A(t)}.
\end{equation}
In Tab. \ref{Tab grav} we report the maximal deviation
defined as $\max \left\{f_{\alpha_c=0}(t) \right\}$
occurring in each curve. In the same way, in Tab. \ref{Tab refr}, we have
collected the maximal fractional deviation of the curves without
cloud refraction $\alpha_r$ with respect to the complete ones.
Tab. \ref{Tab abs} contains the maximal fractional absorption. All
light curves were computed assuming an impact parameter $b=0.2 R_E$, $M=1
M_\odot$ and with $\lambda=4400$ \AA.

With these tables it is easy to distinguish the regimes where the
different phenomena become important, keeping the gravitational
lensing of the central object as reference.

Cloud gravitational lensing can be neglected when the cloud mass
is below a few hundredth of the central mass. Its effects become
heavier for higher cloud masses, maximally for radii of the order
$R_c/R_E \sim 1.5$.

Refraction is always of the same order as cloud gravitational
lensing, dominating only for small radii. It is not a surprise
that the sum of the fractional deviations due to refraction and
gravitational lensing is sometimes greater than one. In fact,
their action is by no way additive, since they simultaneously
concur in the shift of the images and modify the magnification
formula in a non-linear way.

All ranges of the cloud parameters where the refraction
contribution is non-negligible are also affected by a large
absorption. High masses completely block the light as we can read
in the last row of Tab. \ref{Tab abs}. Cloud masses around one
tenth of the central mass still show good positive deviations
coming from light bending, maintaining a moderate absorption. It
is in this mass range that PLAGs have the highest probability of
being detected.

\subsection{Chromaticity}
\begin{figure}
\resizebox{\hsize}{!}{\includegraphics{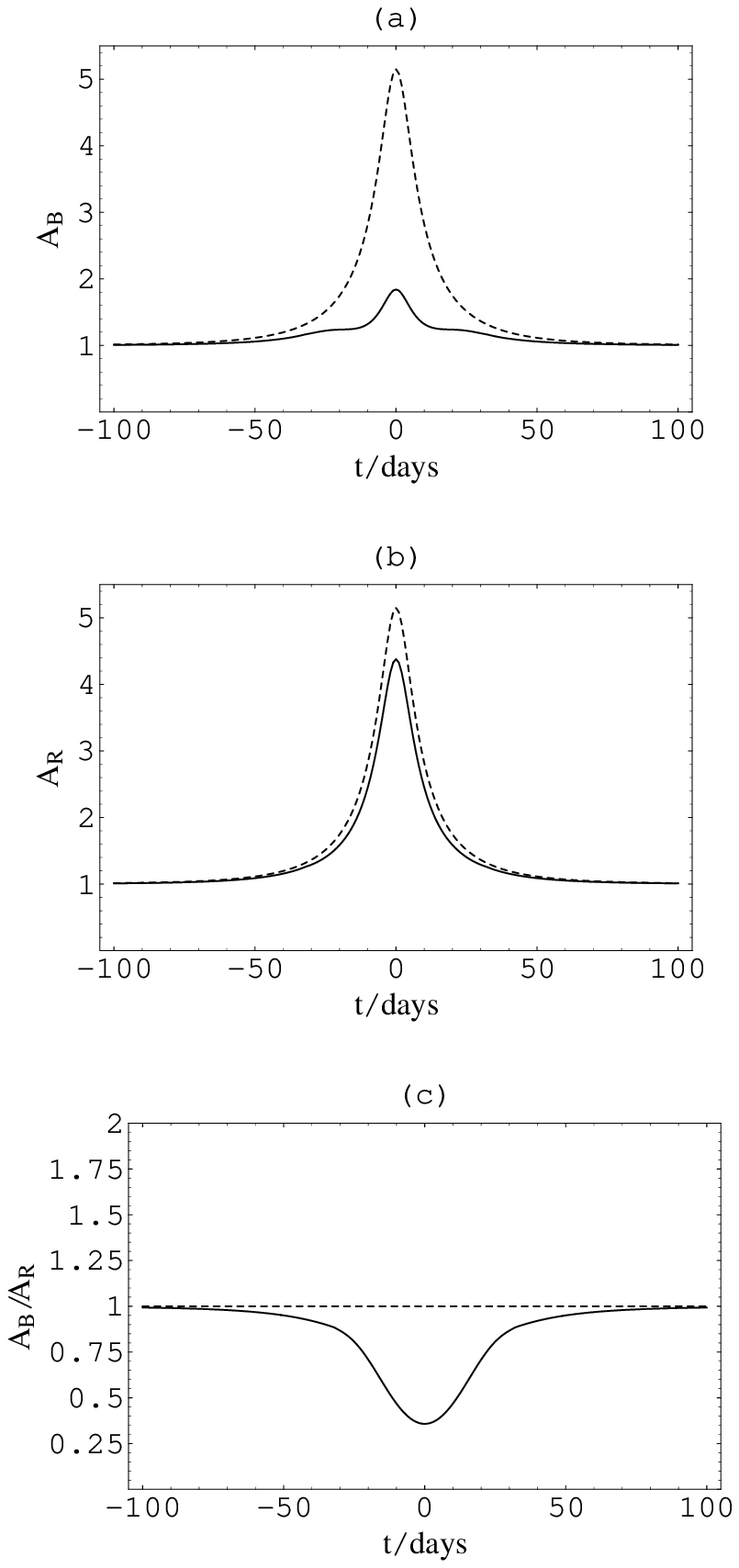}}
 \caption{Chromaticity of the PLAG microlensing event.
(a) $\lambda=4400$\AA (blue). (b) $\lambda=7000$\AA (red). The
dotted lines are again the classical Paczy\'{n}ski light curves.
(c) Ratio between the two curves.}
 \label{chromaticity}
\end{figure}

Both refraction and absorption have a moderate dependence on the
wavelength of the incoming wave. This means that a strong
chromaticity should be expected for PLAG events, mostly caused by
a selective absorption. In a minor extent, blending and refraction
may slightly modify the color. Given the $\lambda^{-4}$ dependence
of the absorption coefficient, Eq.(\ref{absorption coefficent}),
the light filtered through the PLAG would be strongly reddened, as
it is evident from Fig.\ref{chromaticity}, where we reported the
light curve of the example treated in Fig.\ref{Fig Sample Curves}c
at different wavelenghts: $\lambda=4400$ \AA \, (blue),
Fig.\ref{chromaticity}a, and $\lambda=7000$ \AA \, (red),
Fig.\ref{chromaticity}b.
For this reason, the achromaticity test of classical microlensing
surveys would automatically reject PLAG events and should be
relaxed in order to find them. Moreover, the detection of a PLAG
would be helped choosing filters operating in opposite bands of
the spectrum so as to put in evidence the real nature of the
chromaticity in an optimal way.

\subsection{Boundary discontinuities}

In expanding shells produced by old stars, the density profile is
generally characterized by a steep boundary, where the density
suddenly changes from zero to a value significant for refraction
and absorption effects. Of course, this is not the case for cold
dark PLAGs on which we are mainly focused in this article, which
are supposed to be stable or long--lived systems; but a wide range
of alternative PLAG candidates (see Sect. 4) require an
appropriate description of boundary effects.

For this reason, in this subsection we check whether interesting
observable effects could be expected when one of the images enters
an expanding steep-front cloud.

First of all, let us try a perturbative resolution of the lens
equation for $x \simeq R_c$. We let
\begin{equation}
y=R_c-\frac{R_E^2}{R_c}-\delta y \label{y expansion}
\end{equation}
and
\begin{equation}
x=R_c-\varepsilon. \label{x expansion}
\end{equation}

In the neighborhood of $R_c$, $\rho_\beta(r)$ falls to zero as
\begin{equation}
\rho_\beta(r)=f(r) \left( 1-\frac{r}{R_c} \right)^\beta \label{rho
expansion}
\end{equation}
with $f(r)$ a regular non vanishing function at $r=R_c$. This
expansion can be performed in general, not only for the sample
family studied so far, but also for a generic density profile.

The projected density will go as
\begin{equation}
\Sigma_\beta \simeq 2\sqrt{2}f(R_c)\frac{\varepsilon^
{\beta+\frac{1}{2}}}{R_c^{\beta-\frac{1}{2}}}.
\end{equation}

Expanding all deviation angles to first order in $\varepsilon$, we
get
\begin{eqnarray}
\alpha_0 &\simeq& \frac{R_E^2}{R_c}\frac{M}{M+M_c}\left( 1+
\frac{\varepsilon}{R_c} \right) \\ %
\alpha_c &\simeq& \frac{R_E^2}{R_c}\frac{M_c}{M+M_c}\left( 1+
\frac{\varepsilon}{R_c} \right) \\ %
\alpha_r &\simeq& \alpha_\lambda 2\sqrt{2}f(R_c)\left(
\beta+\frac{1}{2} \right) \frac{\varepsilon^
{\beta-\frac{1}{2}}}{R_c^{\beta-\frac{1}{2}}}=A\varepsilon^
{\beta-\frac{1}{2}},
\end{eqnarray}
which can be taken seriously only for $\beta \geq \frac{1}{2}$.
For lower values, a cut-off from higher order terms should be
expected.

The perturbed lens equation reads
\begin{equation}
\delta y=\varepsilon+A\varepsilon^
{\beta-\frac{1}{2}}+\frac{R_E^2}{R_c^2}\varepsilon \label{Pert LE}
\end{equation}
and the magnification for an image close to the boundary is
\begin{eqnarray}
\mu&&=\left[ 1- \frac{R_E^2}{R_c^2}
\left(1+\frac{2\varepsilon}{R_c} \right)-\frac{A}{R_c}
\varepsilon^{\beta-\frac{1}{2}} \right]^{-1} \times \nonumber \\
&&\times \left[ 1+
\frac{R_E^2}{R_c^2}\left(1+\frac{2\varepsilon}{R_c} \right)+\left(
\beta -\frac{1}{2} \right)A \varepsilon^{\beta-\frac{3}{2}}
\right]^{-1} \label{Expanded Mag}
\end{eqnarray}

\begin{figure}
\resizebox{\hsize}{!}{\includegraphics{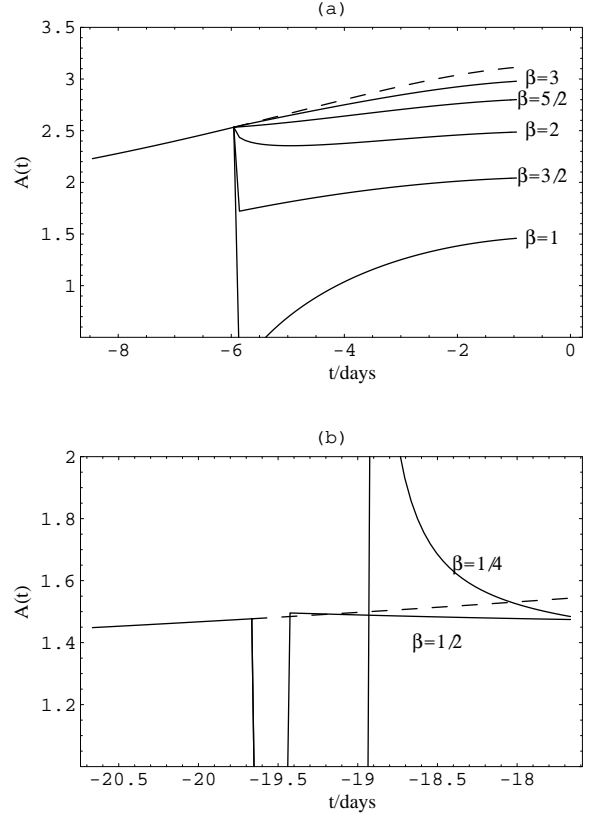}}
 \caption{Details of light curves at the entrance of an image into the cloud.
  The dashed curve is the classical Paczy\'{n}ski.
 (a) Plots for $\beta=3,5/2,2,3/2,1$. (b) Plots for $\beta=1/2,1/4$. }
 \label{Fig Disc}
\end{figure}

Refraction yields $\beta$-dependent terms to the lens equation and
to the magnification. It is easy to analyze the boundary effects
that may arise for low values of $\beta$. In Fig. \ref{Fig Disc},
the light curve is shown for different values of $\beta$ at the
time of the entrance of an image inside the cloud.

According to the value of $\beta$, we can have different cases.

\begin{enumerate}
\item{$\beta>5/2$: The refraction terms introduce subdominant
 corrections. The magnification is a continuous function.
 }
\item{$\beta=5/2$: The first
derivative of the magnification has a discontinuity, being higher
for $r<R_c$. The light curve suddenly decreases its derivative
when an image enters the cloud.
 }
 \item{$3/2<\beta<5/2$: The refraction term becomes dominant in the magnification. The
 derivative of the magnification goes to infinity in $R_c$ and the
 light curve has a continuous depression when an image comes into the
 cloud.
 }
 \item{$\beta=3/2$: The refraction terms in the magnification do
 not vanish on the border. The light curve has a finite discontinuity
 }
 \item{$1/2<\beta<3/2$: The refraction term in the second square
 brackets of Eq.(\ref{Expanded Mag})
 diverges and the magnification falls off to zero. Then it rises
 from zero to the value it would have without refraction.
 }
\item{$\beta=1/2$: The lens equation has no solution until $\delta y$ becomes
 greater than $A$. This fact provokes a hole in the light curve at the entrance of the image
 into the cloud, illustrated in Fig. \ref{Fig Disc}b for our sample model with $\beta=1/2$.
 }
\item{$0 \leq \beta<1/2$: Again we have a hole in the light curve.
 Increasing $\delta y$ further, two values of $\varepsilon$ solve
 the equation. One approaches the original linear relation,
 behaving as the continuation of the original image. The other
 approaches zero, i.e. an additional image is formed very close to
 the border of the cloud.

 So, when an image enters the cloud, after a short interval where
 it temporarily disappears, it comes out doubled, with the
 additional image staying close to the border. The formation of
 this pair happens on a radial critical curve\footnote{We recall that
 the critical curves are defined by the vanishing of the Jacobian determinant of
 the lens application. In axisymmetric lenses
 a critical curve is tangential if $y(x)=0$, while is radial if
 $\frac{\partial y}{\partial x}=0$.
 } and the magnification
 diverges at the formation of the two images.
}
\end{enumerate}

These boundary features produced by refraction are a very
interesting tool to investigate the shape and the nature of the
cloud; but they last few hours and would be hard to detect. It
must be recalled that for such steep variations the finite size of
the source intervenes to smooth peaks and holes. Looking at the
their typical time scales, these features would not be more
elusive than planetary microlensing, after all. For very sharp
density profiles, higher order corrections in the refraction
formula should be taken into account for a correct description at
very low $\beta$.

\subsection{Additional critical curves}

In the previous subsection, we have seen that the structure of the
Jacobian can be predicted once we know the expansion of the
density distribution near the boundary (\ref{rho expansion}). In
this way, we have found that refraction would produce radial
critical curves very close to the boundary in densities with very
steep fronts\footnote{Actually, for $\beta<1/2$ and $R_c>R_E$,
between the additional radial critical curve and $R_c$ there is
also a tangential critical curve, but this feature is certainly a
product of the small angle approximation in $\alpha_r$, since it
comes for very high refraction angles. }. Are these the only
possible additional critical curves?

In isolated clouds, without the central object, critical curves
and caustics can be produced when the density is sufficiently
high, as discussed by Draine (1998). However, when a central mass
is present and the cloud mass is non--dominant with respect to the
central mass, it is far more difficult to create critical curves
as we shall discuss in the following.

To have a tangential critical curve, we need the RHS of the lens
equation (\ref{lens equation}) to vanish. Usually the Einstein
ring is the only critical curve of this class. It is difficult to
have additional curves inside the Einstein ring, because we would
need a very highly negative refraction angle to compensate the
other terms. This kind of refraction angle could only come out in
shell densities with a very steep internal front. Additional
tangential critical curves can be created outside of the Einstein
ring with very high positive refraction angles. Such angles are
beyond the reach of a linear theory and higher order terms would
cut off the refraction angle excluding the formation of tangential
critical curves.

The situation is similar for radial critical curves. Here we have
to kill the derivative of the RHS of (\ref{lens equation}), i.e.
we have to satisfy the equation
\begin{eqnarray}
&1+&\frac{D_{ol}D_{ls}}{D_{os}}[ \alpha_\lambda \Sigma''(x)+
\nonumber \\ && +\frac{4G}{c^2} \left(\frac{M+m(x)}{x^2}-2\pi
\Sigma(x) \right) ]=0. \label{Radial Curves}
\end{eqnarray}

Without the refraction term, this condition reduces to the usual
gravitational lensing one (Subramanian \& Cowling 1986).
Refraction helps to reach this condition but it becomes effective
at high densities. In the range of masses we are interested into,
additional images would be formed only in optically thick clouds
so that no signature of their existence would be visible on the
light curves. On the other hand, decreasing the central mass and
the mass of the cloud correspondingly, the role of refraction
becomes prominent while the cloud becomes more and more
transparent. Finally, going to the regime studied by Draine (1998)
(no central mass and $M_c \sim 10^{-3} M_\odot$), caustic--events
start to be definitely observable. However, in PLAGs, the presence
of a relevant central mass pushes the threshold given by Eq.
(\ref{Radial Curves}) far beyond a high density regime, where the
event just looks like a complete occultation (Fig. \ref{Fig Sample
Curves}e).

Again, only in steep fronts it is possible to have radial critical
curves as we saw in Sect. 3.4 for $\beta<1/2$.

\section{Physical Nature of PLAGs}

In this section we examine the possible PLAG candidates, on the
basis of the phenomenology risen up to now. We have seen what the
microlensing light curves should look like and we have established
an optimal range for the parameters.

As regards cold PLAGs in the halo, i.e. MACHOs associated to gas
clouds, they would be ideal candidates, since their size could be
about of the same order of the Einstein radius. The existence of
such objects in the halo would have very interesting consequences
in our knowledge about the gas distribution in the halo. It could
be argued that a large fraction of cold baryonic mass in the halo
can be found as gas surrounding MACHOs, since the small masses of
these objects would favour the stability of the clouds. If this is
the case, the density of PLAGs may be comparable to ordinary
MACHOs, and several PLAG events per year can be expected by
observing towards the Magellanic Clouds.

Besides cold dark PLAGs in the halo, which surely represent the
most interesting possibility, in this Section we will examine
known types of nebulae to look for additional PLAG candidates.

Circumstellar disks are commonly found at IR wavelenghts around
young stars and around low mass and intermediate stars that are in
their evolutionary stage from red giants to white dwarfs.

Proto-stars could be good candidates. In fact, they are isolated
interstellar clouds undergoing gravitational collapse. The clouds
are actually flattened to disks with circular or elliptical
projected shape according to the tilt angle with respect to the
line of sight. Proto-stars can be found in star-forming regions,
usually located in the disk and in the spiral arms of the Milky
Way. Of course, to observe light curves as those shown in
Fig.\ref{Fig Sample Curves}, we are mainly interested in
Proto-stars with $1<M/M_{c}<10$ and $R_{c}\sim 1-10^{2}$ AU.

In the last decade, a good number of Proto-stars have been
detected in our Galaxy (see for instance Beckwith et al. 1990,
Hillenbrand et al. 1992, McCaugherean, O'Dell 1996, Schneider et
al. 1999, Padgett et al. 1999, Tuthill et al. 2001) and, very
recently, the HST found also a very compact star-forming region in
LMC (Heydary-Malayeri et al. 2001). Twin proto-planetary disks,
each of them with $R_{c}\approx 20$ AU, have also been observed
(Giovanardi et al. 2000), suggesting the possibility of binary
PLAG events.

Other good candidates could be the proto-planetary nebulae (PPNe).
They represent a transitory stage, between the asymptotic giant
branch (AGB) and planetary nebulae phases, in the evolution of
intermediate and low-mass stars. This stage is characterized by an
expanding circumstellar envelope of gas surrounding a central star
(CPPN). In our phenomenological description, we can consider the
PPNe as static systems, because the expansion velocity of the
circumstellar envelope is $v_{e}\approx 25 \, km sec^{-1}$
(Osterbrock 1989). The typical mass of the CPPN is $M\approx 0.6
\, M_{\odot}$ and the mass of the circumstellar envelope is of the
same order. Since they are expanding systems, it is certainly
possible to observe PPNe with $R_{c}\sim R_{E}$. PPNe were only
discovered and studied in the last twelve years, mostly thanks to
{\it IRAS} and {\it HST} (Ueta et al. 2000). For a good overview
about these objects, see Kwok 2000.

As well as in our galaxy, where we know roughly $10^{3}$ planetary
nebulae (PNe), intracluster PNe (IPNe) have also been detected
thanks to the surveys towards the Virgo cluster (Mendez et al.
1997), M87 and its surrounding halo (Ciardulo et al. 1998,
Feldemeier et al. 1998).

Besides Proto-stars and PPNe, we can consider other kinds of
astrophysical objects that are associated with gas clouds. For
instance, recently the VLT discovered a small cone-shaped nebula
near the the old neutron star RX J1856.5-3754. This nebula extends
for roughly $60$ AU (van Kerkwijk 2001), and the neutron star is
located very close to the top of the cone. Similar bow wave nebula
have been found around other fast-moving supernova remnant, like
IC 443 (Olbert et al. 2001), radio pulsars, like PSR B1706-44 (PSR
J1709-4428) and PSR B1643-43 (PSR J1646-4346) (E. Giacani et al.
2001), nova system, like BZ Cam (Prinja et al. 2000), and could be
good PLAG candidates.

All these objects are located mainly in the bulge, in the disk and
in spiral arms of galaxies. So, good targets for the search of
this kind of PLAGs are the Galactic bulge and rich star fields of
the Milky Way spiral arms. However, the monitoring of other
galaxies, like LMC or M31, should give prominence to self-lensing
PLAG events. It must be said that Protostars, PPNe and other
exotic objects are not very common in the galaxy, so that their
total optical depth in observations towards the bulge barely
reaches $10^{-11}$ which is 5 orders of magnitude lower than disk
star optical depth.

In a cosmological context, following the idea to find
extra-galactic MACHOs by monitoring quasars behind the Virgo
cluster of galaxies (Tadros et al. 1998), or by monitoring M87
with the pixel lensing technique (Gould 1995), the search of
intracluster PLAG events could give interesting results. However,
considering the actual estimates of the abundance of intracluster
nebulae (Ciardullo et al. 1998), the optical depth towards M87 is
around $10^{-12}$, which is still very low.

\section{Summary}
We studied gravitational microlensing due to point-like sources
surrounded by diffuse gas clouds. Provided the size of the gas
cloud is about of the order of the corresponding Einstein radius,
we find that the presence of the gas cloud, via absorption or
refraction effects, can substantially modify the classical
Paczy\'nski light curve and induce a strong chromaticity. The
observation of these effects might not be easy, since they require
a precise photometry and a continuos monitoring, as done for
instance by the Planet and GMAN collaborations to find planets.
Otherwise, if one looks for long duration events, the observations
should last several years, in order to identify longer events.

We have hypothesized that cold dark PLAGs may be present in
galactic halos and we have mainly focused on the determination of
the characteristics they should have to be observable. If the
existence of PLAGs in the halo is proved, then the recent
estimates of baryonic dark matter should be revised to take into
account the presence of gas clouds in the halo. Disk PLAG events,
caused by known nebulae, or events caused by intracluster nebulae
should be very rare, if the estimates of their abundances are
correct.

Due to the chromaticity, the present used selection criteria for
microlensing reject PLAG events. It would thus be interesting to
check whether already in the existing data there are PLAG
microlensing events. We believe that a search for PLAG type events
is necessary to complete the picture about baryonic dark matter in
the halo.

\begin{acknowledgements}
V.B. and G.S. were supported by fund ex $60\%$ DPR 382/80 and
F.S.E. of European Community. Ph.J. and L.M. acknowledge support
from the Swiss National Science Foundation.
\end{acknowledgements}

\end{document}